\documentclass[aip,jcp,preprint,amsmath,amssymb]{revtex4-1}

\draft 

\usepackage{natmove} 
\usepackage{graphicx}
\usepackage{dcolumn}
\usepackage{bm}
\usepackage{color} 

\begin{document}


\title{Exciton localization in tubular molecular aggregates: size effects and optical response} 



\author{Anna S. Bondarenko}
\author{Thomas L. C. Jansen}
\author{Jasper Knoester}
\email[]{j.knoester@rug.nl}
\affiliation{$^1$University of Groningen, Zernike Institute for Advanced Materials, Nijenborgh 4, 9747 AG Groningen, The Netherlands}

\date{\today}

\begin{abstract}
We study the exciton localization and resulting optical response 
for disordered tubular aggregates of optically active molecules. 
It has been shown previously that such tubular structures allow 
for excitons delocalized over more than a thousand molecules, 
owing to the combined effects of long-range dipole-dipole interactions 
and the higher-dimensional (not truly one-dimensional) nature of 
the aggregate. Such large delocalization sizes prompt the question 
to what extent in experimental systems the delocalization may still 
be determined by the aggregate size (diameter and length) and 
how this affects the aggregate's optical response and dynamics. 
We perform a systematic study of the size effects on the localization 
properties, using numerical simulations of the exciton states in 
a cylindrical model structure inspired by the previously derived 
geometry of a cylindrical aggregate of cyanine dye molecules (C8S3). 
To characterize the exciton localization, we calculate the participation 
ratio and the autocorrelation function of the exciton wave function. 
Also, we calculate the density of states and absorption spectrum. 
We find strong effects of the tube's radius on the localization and 
optical properties in the range of parameters relevant to experiment. 
In addition, surprisingly, we find that even for tubes as long as 750~nm, 
the localization size is limited by the tube's length for disorder values 
that are relevant to experimental circumstances, while observable 
effects of the tube's length in the absorption spectrum still occur for 
tube lengths up to about 150 nm. The latter may explain changes 
in the optical spectra observed during the aging process of 
bromine-substituted C8S3 aggregates. For weak disorder, the exciton 
wave functions exhibit a scattered, fractal-like nature, similar 
to the quasi-particles in two-dimensional disordered systems.
\end{abstract}


\maketitle 

\section{Introduction}

Self-assembled aggregates of molecules with strong optical transitions 
have been studied abundantly for more than 80 years now \cite{Jelley.1936, 
Jelley.1937, Scheibe.1937}. 
The close packing of molecules within such aggregates gives rise to collective
optically allowed excited states, Frenkel excitons, that are shared by a number
of molecules and that give rise to interesting optical phenomena. Examples are
exchange narrowing of spectral lineshapes \cite{Knapp.ChemPhys.1984, 
Malyshev.motional.1999}, collective spontaneous emission \cite{DeBoer.esr.1990, 
Fidder.esr.1990}, a Pauli-exclusion gap measured in pump-probe spectroscopy 
\cite{Fidder.pump-probe.1993, Juzeliunas.linear.agg.1988}, and enhanced nonlinear 
optical properties \cite{Spano.Chi3.1989, Knoester.chi3.1993}. Typically, such 
aggregates also exhibit fast excitation energy transport, reflected, for instance, 
in very high exciton-exciton annihilation efficiencies \cite{Scheblykin.thiats.2001}. 
Synthetic dye aggregates consisting of many thousands of molecules, in particular 
those prepared from cyanine dye molecules, have played a crucial role in the
development of color photography and xerography \cite{Tani.photo.1995, Herz.1977, 
Mees.1942}. On the other hand, natural aggregates consisting of optically active 
biomolecules also have received much attention lately, in particular in 
the context of light-harvesting antenna complexes in the photosynthetic systems 
of bacteria, algae, and higher plants \cite{Scholes.NatChem.2011, Orf.Photo.2013, 
Wurtner.NatChem.2016, Jang.Mennucci.2018}. Such aggregates, mainly consisting 
of (bacterio)chlorophyll molecules, usually stabilized in a protein scaffold, have 
the purpose of absorbing the energy of the sunlight, thereby converting it into 
an electronic excitation, which subsequently is transported with high efficiency 
(quantum efficiencies over 90\%) to the photosynthetic reaction center to trigger 
the first step in the photochemical reaction. The extent to which delocalized and 
quantum coherent excitons play a role in natural antenna systems has been a topic 
of much interest during the past 20 years \cite{Oijen.LH2.Sci.1999, Jang.B850.JPCB.2001,
Reineker.Jlum.2003, Engel.Nature.2007, Wu.Cao.JCP.2012, Cleary.2013.NJP.15.125030,
Tempelaar.FMO.2014, Scholes.Nature.2017, Duan.PNAS.2017, 
Jang.Mennucci.2018, Tempelaar.FMO.NatChem.2018, Celardo.NJP.2019, 
Celardo.microtubules.2019}.

The role of collective effects in the optical response and excited state dynamics 
of molecular aggregates depends on how many molecules share an excitation, 
a quantity known as the exciton delocalization size. In ideal, nicely ordered 
aggregates of identical molecules, in principle the excitations are shared by 
all molecules. In practice, however, disorder in the transition energies of individual 
molecules imposed by an inhomogeneous host medium and disorder in the excitation 
transfer interactions between molecules, resulting from structural fluctuations, limit 
the exciton delocalization to much smaller numbers. In the prototypical aggregates 
of the synthetic dye molecule pseudo-isocyanine (PIC), the delocalization size at 
low temperatures is in the order of 50-70 molecules \cite{Fidder.JCP.1991, 
Fidder.pump-probe.1993}, which is large enough to see strong collective effects, 
but still considerably smaller than the many thousands of molecules that make up 
these aggregates. The strong localization effect results from the one-dimensional 
character of PIC aggregates.

During the past 15 years, a large number of tubular molecular aggregates 
have been studied, both synthetic~\cite{Spitz.ChemPhys.2002, Spitz.2006, 
Gandini.2003.small,Vlaming.2009.exciton, Pawlik.JCPB.1997, Didraga.JPCB.2004, 
Eisele.NatChem.2012, Clark.JPCC.2013, Sperling.2010.excitons,
Abramavicius.PRL.2012, Yuen.ACSnano.2014, Lang.2005.optical, Sengupta.2012, 
Berlepsch.Langmuir.2013, Sorokin.2010.control, Womick.2009.probing, 
Womick.2009.correlated,Doria.ACSnano.2018,Caram.NanoLett.2016,Kriete.NatComm.2019}, 
semi-synthetic \cite{Sengupta.2012, Tenzin.ZnChl.2019} 
and natural \cite{Holzwarth.1994,Ganapathy.PNAS.2009, Tian.JACS.2011, 
Gunther.JPCB.2016} ones. These systems typically have diameters in the order 
of 10 nm and lengths of 100 nanometer up to microns. This renders them 
quasi-one-dimensional systems from a geometrical point of view, that might, 
for instance, serve as excitation energy transport wires. However, it has been 
shown that the extra dimension (the tube's circumference) in combination with 
the long-range (dipolar) intermolecular excitation transfer interactions leads to 
much weaker exciton localization than in truly one-dimensional systems 
\cite{Bloemsma.PRL.2015}. This explains experiments on a variety of tubular 
aggregates, demonstrating strong dependence of the optical properties on 
the polarization direction of the absorbed or emitted light relative to the tube's 
axis \cite{Spitz.ChemPhys.2002, Spitz.2006, Didraga.JPCB.2004, Clark.JPCC.2013, 
Gunther.JPCB.2016}. In fact, the delocalization size in tubular aggregates of 
the dye C8S3 was estimated to be in the order of a thousand molecules, even 
in the optically dominant energy region near the exciton band edge, where 
localization properties are strongest \cite{Bloemsma.PRL.2015}.

The large exciton delocalization sizes in tubular aggregates are of direct 
relevance to their optical and excitation transport properties. For instance, 
it has been shown that the exciton diffusion constant in tubular model  
aggregates is a universal function of the ratio of the exciton localization 
length and the cylinder's circumference \cite{Chuang.PRL.2016}. 
This becomes all the more interesting, because recently some degree of 
control of the radius of tubular aggregate of cyanine molecules has been 
reported\cite{Kriete.JPCL.2017}. Moreover, given the large delocalization 
sizes obtained in numerical simulations, the question arises to what extent 
the system size still plays a role in their value, both in the calculation and 
in experiment. Thus far, a systematic study of size effects in the localization 
properties, and hence the optical properties, has not been performed.

In this work, we numerically investigate the dependence of the exciton
localization properties and absorption spectrum on both the radius and 
the length of tubular molecular aggregates. We employ a Frenkel exciton 
model with Gaussian site disorder on an experimentally relevant tubular 
aggregate structure. The findings confirm that under experimental conditions, 
it is possible that the delocalization is not solely determined by the ratio of 
the strength of the disorder and the width of the exciton band, but also by 
the aggregate size. We also show that this does not imply that the excitons 
are spread over the entire system in the same way as the excitons in 
a homogeneous tubular aggregate are; rather the wave functions seem 
to be spread in a highly irregular way, resembling fractal behavior
\cite{Schreiber.1991.multifractal}.

The outline of this paper is as follows. In Sec.~\ref{Deloc:sec:Model}, 
we describe the details of the model used in our study and the approach; 
in particular, we define the various quantities studied in our analysis. 
Next, in Sec.~\ref{Deloc:sec:Results}, we present our results, followed 
by a discussion. Finally, in Sec.~\ref{Deloc:sec:Conclusions}, we conclude.
In the appendix several details are presented that characterize the exciton 
band as a function of system size.

\section{Model and Approach} 
\label{Deloc:sec:Model}

\subsection{Structural model}

Throughout this paper, we use as model system the extended herringbone
model introduced in Ref.~\citenum{Eisele.NatChem.2012} to describe the inner 
wall of the frequently studied double-walled tubular molecular aggregates of 
the dye C8S3 (3,3$^\prime$-bis(2-sulfo{\-}propyl)-5,5$^\prime$,6,6$^\prime$-tetrachloro-1,1$^\prime$-dioctyl{\-}benzimida{\-}carbo{\-}cyanine). 
This model describes a single-walled tubular aggregate with two identical 
molecules per unit cell, which only differ from each other by their position 
in the unit cell and their orientation in the local frame of the tube. For the 
purpose of describing the optical properties, all molecules are considered 
two-level systems with an optical transition dipole between the ground state 
and the excited state that is fixed to the molecular frame. The model may 
be considered as a perpendicular stack of $N_1$ equidistant rings, separated 
by a distance $h$, where on each ring the positions of $N_2$ equidistant unit 
cells are located. Neighboring rings are rotated relative to each other over 
a helical angle $\gamma$.

The above described packing is realized by wrapping a planar two-dimensional
lattice with two molecules per unit cell (which are tilted out of the plane) on
a cylindrical surface. This wrapping is fully dictated by the length and
direction of the vector over which the lattice is rolled; the length of
the rolling vector gives the circumference of the cylinder (and hence dictates
the radius of the tube). The parameter $h$ only depends on the orientation of
the rolling vector and the lattice constants, while both $N_2$ and $ \gamma$
also depend on its length.  Throughout this work, the orientation of the rolling
vector was kept fixed and equal to the one used for the inner wall in Ref.
~\citenum{Eisele.NatChem.2012} to fit the experimental spectrum of C8S3 tubes, 
leading to a fixed value of $h=0.2956$~nm. In order to allow us to investigate 
the dependence of the tube's localization and optical properties on its radius, 
the length of the rolling vector was varied. This means that only a discrete set 
of radii can be considered, because after wrapping the two-dimensional lattice 
on the cylinder, the molecule where the rolling vector starts should coincide with 
the one where it ends (seamless wrapping). For further details of creating the structural
model, in particular the lattice parameters, the tilt angles, and the orientation of 
the rolling vector, we refer to Refs.~\citenum{Eisele.NatChem.2012} and 
\citenum{Kriete.JPCL.2017}.

The variation of radii considered in our calculations is such that $N_2$ takes
all integer values in the range $N_2 = 1,..., 35$. Within the model considered here, 
the inner tube of the C8S3 aggregates has $N_2=6$, while the inner wall of
the wider bromine-substituted C8S3 
(3,3$^\prime$-bis(2-sulfo{\-}propyl)-5,5$^\prime$,6,6$^\prime$-tetrabromo-1,1$^\prime$-dioctyl{\-}benzimida{\-}carbo{\-}cyanine) aggregates has $N_2=11$, both values that fall inside the
range studied here. When investigating the radius dependence, the length was
kept fixed at $N_1=666$, which agrees with a physical length of 196.9~nm. When
studying the length dependence, the radius was kept fixed at 3.5505~nm, which
agrees with a ring of $N_2=6$ unit cells, the value that applies to the inner wall 
of unsubstituted C8S3. We then considered 11 different values for the length lying between
$N_1=170$ and $N_1=2500$, i.e., a physical length between 50~nm and $740$~nm,
which is an experimentally relevant range. The total number of molecules in the
aggregate thus ranges from 1332 to 46,620 for the smallest and largest radii, respectively.

\subsection{Model Hamiltonian}

The collective optical (charge neutral) excited states of the aggregate are 
described by the Frenkel exciton Hamiltonian, 
			\begin{equation} 
			\label{Hamiltonian}
			H=\sum_{n,m} H_{nm} |n\rangle \langle m| = \sum_n (\omega_0 + \Delta_n)
			|n\rangle \langle n| + \sum_{n \neq m} J_{nm} |n\rangle \langle m|,
			\end{equation} 
where $n$ and $m$ run over all molecules, and $|n \rangle$ denotes 
the state where molecule $n$ is excited and all other molecules are in 
their ground state. Throughout this paper, we use open boundary conditions 
at the top and bottom rings of the cylinder (i.e., the cylinder is not folded into a torus).

The first term in Eq.~(\ref{Hamiltonian}) describes the molecular excited state
energy ($\hbar=1$), where $\omega_0$ gives the mean value, which is taken 
to be 18,868 cm$^{-1}$ \cite{Eisele.NatChem.2012}, and the offset $\Delta_n$ 
describes the energy disorder that gives rise to localization. We will model the 
disorder by randomly choosing the $\Delta_n$ from a Gaussian distribution with 
mean zero and standard deviation $ \sigma$; the disorder offsets on different 
molecules are assumed to be uncorrelated from each other. The second term in 
Eq.~(\ref{Hamiltonian}) describes the intermolecular excitation transfer interactions, 
which are described by extended dipole-dipole interactions between all molecules, 
using $q=0.34e$ and $l=0.7$ nm, respectively, for the point charges and length of 
the vector connecting them~\cite{Eisele.NatChem.2012}. No disorder in the interactions 
$J_{nm}$ is considered in this paper, i.e., we do not take into account structural disorder. 
The interactions $J_{nm}$ promote the delocalization of the excitation over the
aggregate. For the structure considered here, the interactions are strong, owing
to the fact that the C8S3 molecules have a large transition dipole (11.4 Debye)
\cite{Eisele.NatChem.2012} and have small separations between each other. 
Based on this, the strongest four interactions have an absolute value between 
1000 cm$^{-1}$ and 1500 cm$^{-1}$, while the next three largest interactions 
all are in the order of $-500$ cm$^{-1}$.

In order to study the localization and optical properties of the aggregate, we
first numerically diagonalize the Hamitonian for a particular disorder
realization, which provides us with the eigenstate $|q\rangle = \sum_{n}
\varphi_{qn} |n\rangle$ and its energy $ \omega_q$, where $\varphi_{qn}$ and
$\omega_q$ denote the eigenvectors (normalized to unity) and eigenvalues,
respectively, of the matrix $H_{nm}$. From these quantities all properties we
are interested in follow. Specifically, the exciton density of states (DOS) is given by
			\begin{equation}
			\label{DOS}
			\rho(\omega) = \bigg \langle \sum\limits_q \delta (\omega-\omega_q) \bigg \rangle, 
			\end{equation}
where the angular brackets denote an average over disorder realizations. 
Similarly, the absorption spectrum is given by
			\begin{equation}
			\label{absorption}
			A(\omega) = \bigg \langle \sum\limits_q |{\vec e} \cdot {\vec \mu}_q|^2 
			\delta (\omega-\omega_q) \bigg \rangle, 
			\end{equation}
where ${\vec e}$ denotes the electric polarization vector of the light used to
take the spectrum and ${\vec {\mu}}_q = \sum_n \varphi_{qn} {\vec \mu}_n$ is 
the transition dipole between the aggregate's ground state (all molecules in their
ground state) and the exciton state $|q\rangle$, with ${\vec \mu}_n$ denoting
the transition dipole vector of molecule $n$.

\subsection{Wavefunction characterization}

To characterize the exciton localization properties, we study two quantities.
The first one is the inverse participation ratio, defined as \cite{Thouless.1974.PhysRep.13.93, 
Schreiber.1982.JPhysSocJpn.5.1537}
			\begin{equation}
			\mathcal{P}^{-1}(\omega) = \frac{ \bigg \langle \sum\limits_q \delta (\omega-\omega_q) 
			\sum\limits_n |\varphi_{q\mathbf{n}}|^4 \bigg  \rangle } { \rho (\omega) } .
			\label{Deloc:eq:PN}
			\end{equation}
The inverse participation ratio equals 1 for states localized on one molecule
only, while for states that are equally shared by all molecules of the aggregate, 
its value equals $1/N$, where $N$ is the total number of molecules.
Alternatively, the reciprocal of the inverse participation ratio, also known as
the participation ratio, $\mathcal{P}(\omega)$, is generally accepted as a
quantity that characterizes how many molecules take part in (share) the
collective excitations at energy $\omega$. Depending on the disorder strength
and the exciton energy, this value may be anywhere between unity (totally
localized state) and $\alpha N$ (totally delocalized state), where $\alpha$ is a
constant in the order of unity, which depends on whether open or closed boundary 
conditions are used.
In general, the localization properties depend on energy $\omega$, as is made 
explicit in the above notation.

The second quantity that we will use to investigate the extended nature of the
exciton states is the autocorrelation function of the exciton wave
function\cite{Didraga.JCP.2004}, derived from
			\begin{equation}
			C_{ij}(\mathbf{s};\omega) = \frac{N_1}{N_1-|s_1|} 
			\frac{ \bigg \langle \sum\limits_q \sum\limits_{\mathbf{n}} 
			| \varphi_q(\mathbf{n},i) \varphi^*_q(\mathbf{n+s},j) | 
			\delta (\omega-\omega_q)  \bigg  \rangle }{\rho (\omega)},
			\label{Deloc:eq:ACF}
			\end{equation}		
where a two-dimensional vector notation has been introduced to indicate the
position of the unit cell and $i$ and $j$ can both take on the values 1 and 2 in
order to label the different molecules within the unit cell. Thus, ${\bf n}
=(n_1,n_2)$, with $n_1=1,..., N_1$ labeling the ring on which the unit cell
resides and $n_2 = 1,..., N_2$ indicating its position in the ring. Because
the aggregate has open boundary conditions in the $n_1$ direction, the number 
of terms in the summation over $n_1$ is limited by the value of $s_1$. This would
possibly result in an artificial fast drop of the correlation function with growing $|s_1|$; 
in order to account for this, the correction factor $N_1/(N_1-|s_1|)$ has been added. 
A similar correction is not needed for the $n_2$ summation, as in the ring direction 
periodic boundary conditions are inherently included in the system, always allowing 
for $N_2$ terms in the summation over $n_2$.

$C_{ij}$ defined above is a $2 \times 2$ matrix, whose elements show an overall
similar decay behavior for localized states. In order to just present one
quantity, we have chosen to focus on one specific correlation function defined
through the trace of the matrix:
			\begin{equation}
			\label{Deloc:eq:CF}
			C(\mathbf{s};\omega) = C_{11}(\mathbf{s};\omega) + C_{22}(\mathbf{s};\omega),
			\end{equation}
which has the nice property that its value at the origin is normalized to unity
at all energies: $C(\mathbf{s}={\bf 0};\omega) = 1$.
From the autocorrelation function, we may extract another localization measure 
$N_{\mathrm{corr}} (\omega)$ as a function of energy as the number of molecules 
for which $|C(\mathbf{s};\omega)|>1/e$~\cite{Didraga.JCP.2004}. We also will be 
particularly interested in the autocorrelation function along the tube's axis and define
the correlation length $N_{\parallel,\mathrm{corr}} (\omega)$ as the number of rings 
for which $|C(s_1,s_2=0, \omega)| > 1/e$. $N_{\parallel,\mathrm{corr}}$ is a measure 
of the number of rings over which the exciton wave functions are delocalized.

In the results presented in the next section, the number of disorder 
realizations used to evaluate the disorder average 
$\langle \cdots \rangle$ was taken to be 150.

\section{Results and Discussion} 
\label{Deloc:sec:Results}

\subsection{Absorption spectra}

\begin{figure}[htbp]
  				\centering
  				\includegraphics[width=1\linewidth]{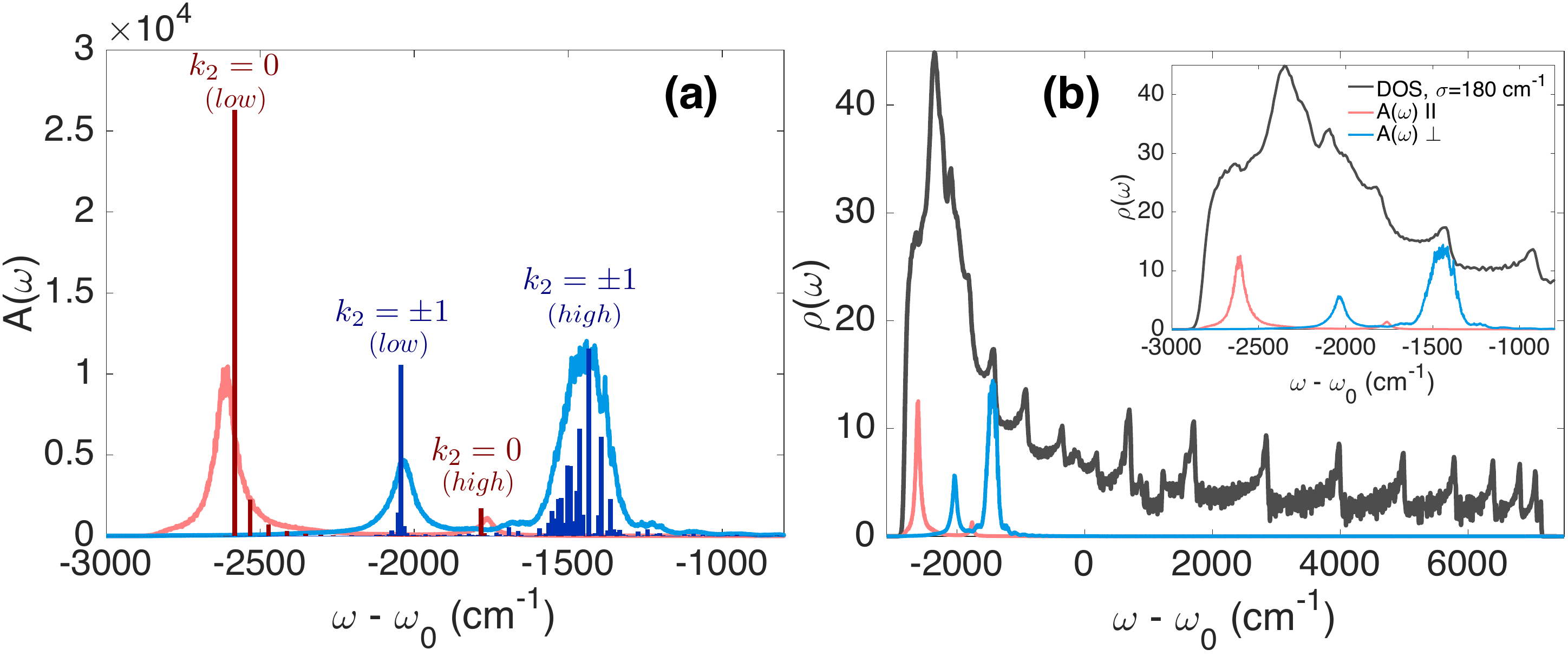}
				\caption{(a) Typical absorption spectrum: homogeneous limit vs disordered case. 
				The stick spectrum of a homogeneous tubular aggregate 
				with $N_1=666$ and $N_2=6$ is shown together with 
				the spectrum in the presence of disorder with $\sigma=180$~cm$^{-1}$. 
				The absorption spectrum has four optical band regions: 
				those polarized parallel (red) and perpendicular (blue) 
				to the tube's axis, each having low- and high-energy Davydov components.
				(b) DOS of the same system with $\sigma=180$~cm$^{-1}$, 
				plotted together with the absorption spectrum depicted in a). 
				The inset shows a magnification of the region of the absorption bands. 
				\label{Deloc:fig:typ_abs}}
				\end{figure}

We start from studying how optical properties depend on the tube's size 
(length and radius) and on the disorder. A typical absorption spectrum of 
the tubular aggregate studied here is shown in Fig.~\ref{Deloc:fig:typ_abs}(a) 
for a structure with length $N_1=666$ and radius $N_2=6$.
The stick spectrum of the homogeneous structure features four \textit{J} bands: 
two bands polarized parallel to the tube's axes (red lines) and two bands 
polarized perpendicular to it (blue lines). These bands originate from the 
selection rules dictated by the cylindrical symmetry of the structure
\cite{Didraga.2002.JPCB.106.11474}. 
%
%
%
%
The eigenstates of the homogeneous tubular aggregate have Bloch 
character in the ring direction characterized by a transverse quantum 
number $k_2$~\cite{Didraga.2002.JPCB.106.11474}. 
The optically allowed states occur in the bands of states with $k_2=0$ 
(polarized parallel to the tube's axis) and those with $k_2=\pm1$ 
(degenerate and polarized perpendicular to the axis). Moreover, the 
lattice with two molecules per unit cell used in this study gives rise to 
a Davydov splitting, resulting in the four optical bands observed in 
Fig.~\ref{Deloc:fig:typ_abs}(a). Our main interest is the low-energy 
Davydov component of the $k_2=0$ band, as this optical band lies 
close to the bottom of the exciton band (see Fig.~\ref{Deloc:fig:typ_abs}(b)) 
and, therefore, has a linewidth that is primarily determined by static disorder.

Disorder gives rise to broadening and an energy shift of the optical 
bands compared to the homogeneous stick spectrum. This is shown 
in Fig.~\ref{Deloc:fig:typ_abs}(a) where light red and light blue lines 
show the spectrum in both polarization directions for tubes with (weak) 
disorder strength given by $\sigma=180$ cm$^{-1}$. The disorder 
strength of 180 cm$^{-1}$ is used, as in the model considered here 
this explains the broadening of the lowest-energy \textit{J} band for bromine-substituted
C8S3 aggregates observed in experiment \cite{Kriete.JPCL.2017}. 
The broadening and energy shift are a result of the breaking of the selection 
rules by the disorder and the resulting mixing of states with different $k_2$ 
values.
The density of states for this system is shown in Fig.~\ref{Deloc:fig:typ_abs}(b) 
together with the absorption spectrum. The exciton band exhibits 
a marked asymmetry around its center as a result of the inclusion 
of long-range interactions \cite{Fidder.JCP.1991}. The lowest-energy 
optical band lies slightly above the lower exciton band edge. 
The energy dependence of the density of states reflects sharp 
peaks due to the one-dimensional sub-bands for different $k_2$ 
values, which persist for the weak disorder value of $\sigma = 180$~cm$^{-1}$.

\begin{figure}[ht]
  				\centering
  				\includegraphics[width=0.6\linewidth]{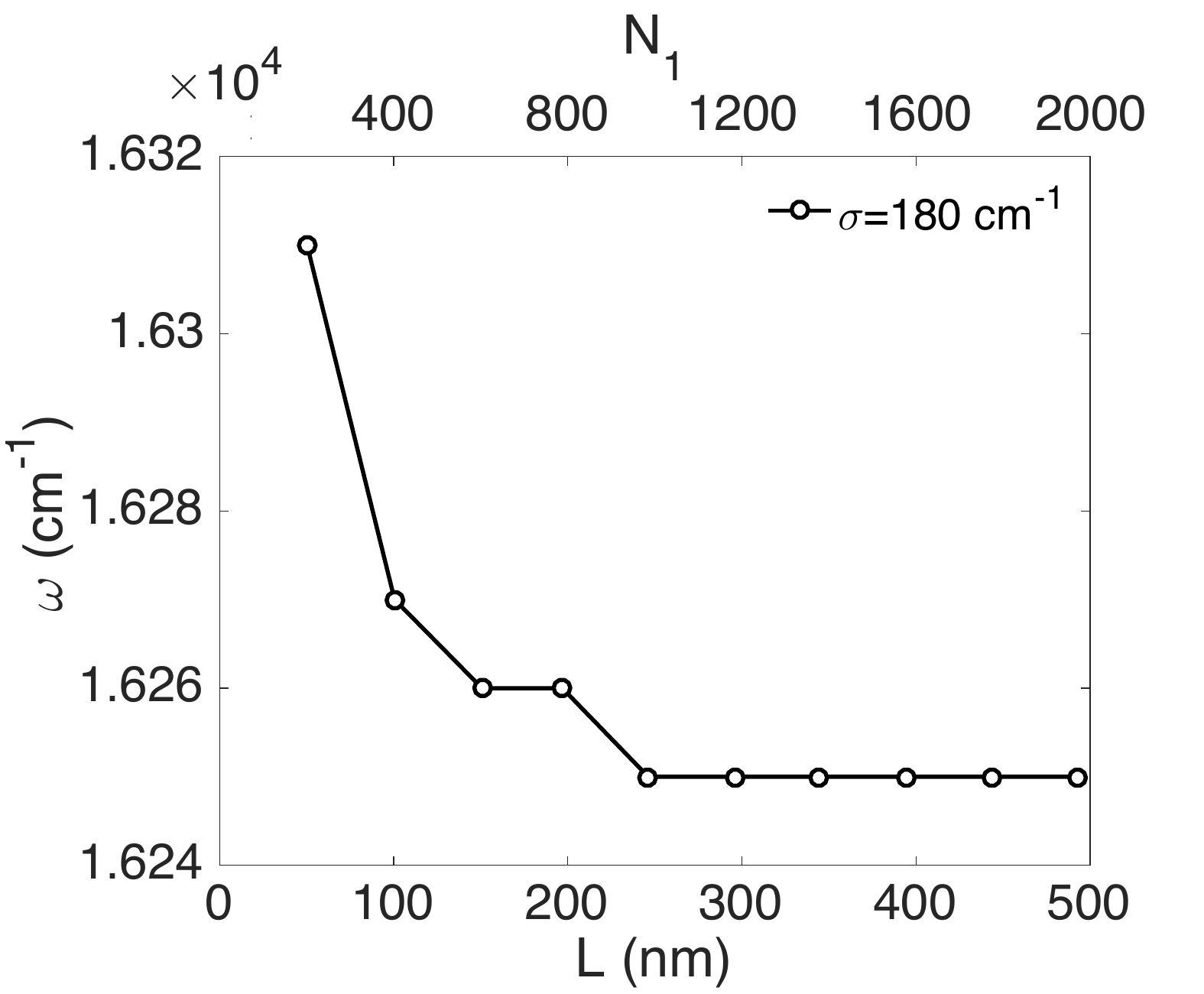}
				\caption{Dependence of the position of the simulated 
				lowest-energy optical band on the tube's length for 
				tubes with $N_2=6$ and $N_1$ ranging from 170 to 
				1666 for $\sigma$=180~cm$^{-1}$.
				\label{Deloc:fig:redshift}}
				\end{figure}

\textbf{Length dependence.}
We first examine the effect of the tube's length on the absorption spectrum. 
Fig.~\ref{Deloc:fig:redshift} displays the calculated position of the lowest-energy 
optical band as a function of the tube's length in the presence of disorder. 
This figure suggests that observable changes in the position of this 
band occur for tubes with lengths up to 150~nm ($N_1=510$). Specifically, 
a red shift of 50~cm$^{-1}$ arises between the tubes with 
$N_1=170$ and $N_1=510$, corresponding to an increase of the length 
from 50~nm to 150~nm. For $N_1>510$, the calculated energy position 
of the optical bands essentially does not change. 
Our theoretical calculations are in qualitative agreement with the experimentally 
observed red shift of the parallel polarized band during the aging process of bromine-substituted
C8S3 aggregate solutions: using cryo-TEM imaging, the tube's length was 
seen to grow while aging.   \cite{Kriete.private}.

\begin{figure}[ht]
  				\centering
  				\includegraphics[width=0.85\linewidth]{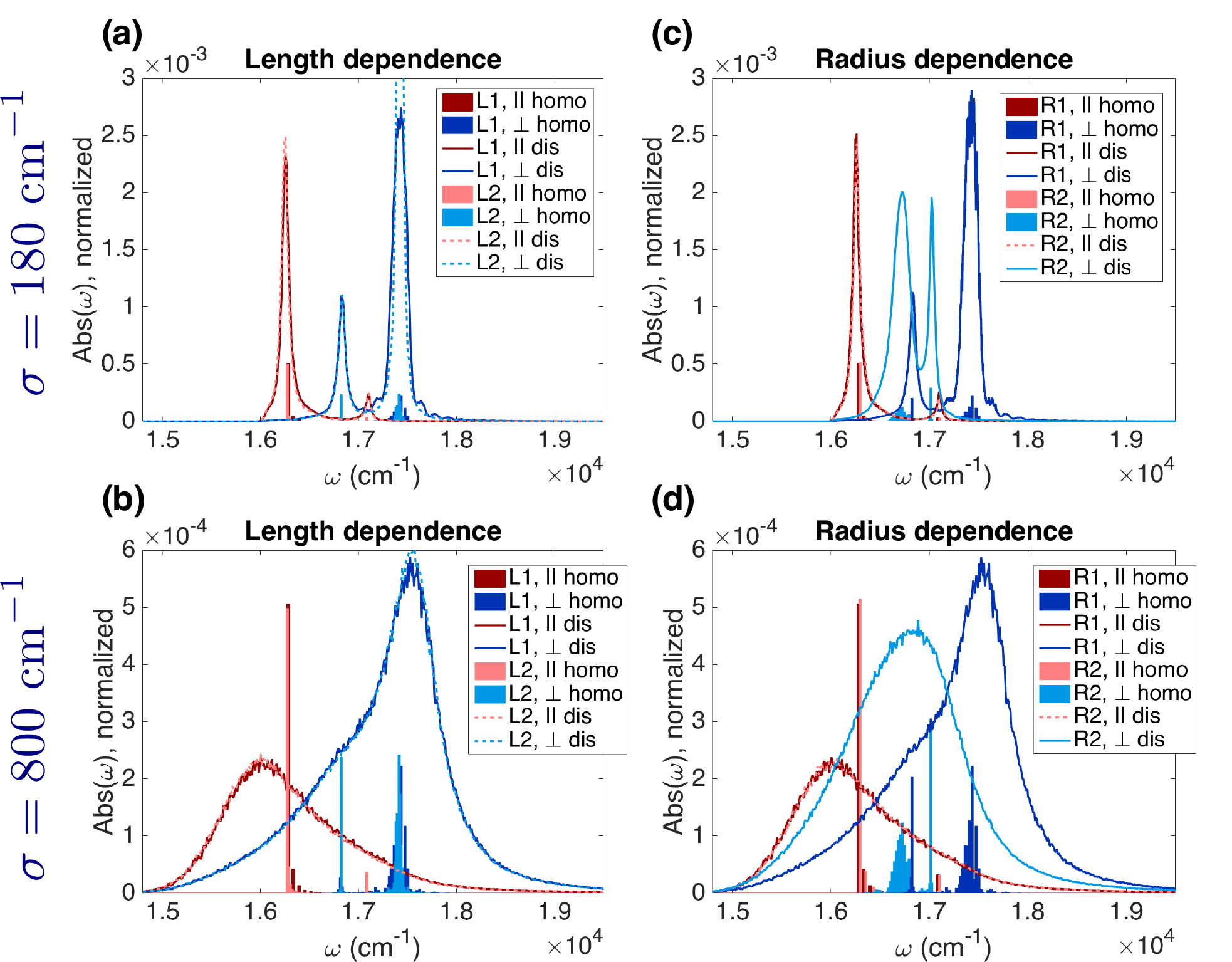}
				\caption{Length and radius dependence of the absorption spectra. 
				Parallel (red colors) and perpendicular (blue colors) 
				polarized bands of the absorption spectra are presented 
				for two different lengths (a,b) and two different radii (c,d) 
				for weak (a,c) and stronger (b,d) disorder.
				The length dependence of the absorption spectra is shown 
				for tubes of fixed radius ($N_2=6$) and smaller length L1, 
				$N_1=833$ (dark red and dark blue) and larger length L2, 
				$N_1=1666$ (light red and light blue). 
				The radius dependence of the absorption spectra is presented 
				for tubes with fixed length ($N_1=666$) and smaller radius R1, 
				$N_2=6$ (dark red and dark blue) and larger radius R2, 
				$N_2=15$ (light red and light blue). 
				The spectra of the disordered systems for both length and 
				radius dependencies, are shown together with the homogeneous 
				stick spectra with the same color scheme. 
				The disordered absorption spectra were normalized to 
				the area under the spectrum. The homogeneous stick 
				spectra were scaled by a factor of 0.02 to facilitate the comparison.
				\label{Deloc:fig:abs}}
				\end{figure}

Next, we examine the overall line shape of the absorption spectra of 
large tubes in the presence of disorder. Figs.~\ref{Deloc:fig:abs}(a) 
and \ref{Deloc:fig:abs}(b) present the absorption spectra of two tubes 
with the same radius ($N_2=6$) but different lengths, decomposed in 
parallel (red) and perpendicular (blue) bands. The two tubes have a length 
of $N_1=833$ (L1) and $N_1=1666$ (L2) and are shown in darker and 
lighter colors, respectively. Absorption spectra of the tubes with weak 
($\sigma=180$~cm$^{-1}$, Fig.~\ref{Deloc:fig:abs}(a)) and strong 
($\sigma=800$~cm$^{-1}$, Fig.~\ref{Deloc:fig:abs}(b)) disorder are 
shown together with the stick spectra in the absence of disorder. 
As can be seen, both for weak and strong disorder the width and 
energy position of parallel and perpendicular bands do not change 
anymore with increasing length in this $N_1$ region. 
This is somewhat surprising in the light of the fact observed later on 
(Sec.~\ref{Deloc:sec:PN}) that at least for $\sigma = 180$~cm$^{-1}$, 
the exciton delocalization size still grows with the tube's length in this region. 
This implies that exchange narrowing of the absorption bands---the effect 
that the absorption band width is inversely proportional to the square root 
of the delocalization size, common for one-dimensional aggregates with 
uncorrelated Gaussian disorder
\cite{Knapp.ChemPhys.1984, Fidder.JCP.1991}---does not occur here. 
This may be related to the special character of the exciton states discussed 
later on (Sec.~\ref{Deloc:sec:fractal}) and the very high density of states 
in the optically relevant region of the spectrum, where bands with different 
$k_2$ values are easily mixed by disorder.

\textbf{Radius dependence.}
Next, we study the effect of the tube's radius on the absorption spectrum. 
Figs.~\ref{Deloc:fig:abs}(c) and \ref{Deloc:fig:abs}(d) present the parallel (red) 
and perpendicular (blue) polarized contributions of the absorption spectra of tubes 
of the same length ($N_1=666$) and different radii; darker and lighter colors 
correspond to tubes with $N_2=6$ (R1) and $N_2=15$ (R2), respectively. 
Increasing the tube's radius gives rise to considerable changes in the 
absorption spectra, which primarily originates from the radius dependence 
of the energy position of the perpendicular polarized optical bands
\cite{Didraga.2002.JPCB.106.11474}. 
The width and position of the lowest-energy optical band is hardly sensitive 
to the radius. This is true for both values of the disorder, $\sigma=180$~cm$^{-1}$ 
(Fig.~\ref{Deloc:fig:abs}(c)) and  $\sigma=800$~cm$^{-1}$ (Fig.~\ref{Deloc:fig:abs}(d)).
The dependence of the energy position of the perpendicular polarized 
band on the tube's radius is the main cause of the changes in the absorption 
spectrum experimentally observed when replacing four chlorine atoms by bromine 
atoms in C8S3 molecules, which leads to larger radii of the self-assembled 
nanotubes.~\cite{Kriete.JPCL.2017}.

\begin{figure}[ht]
  				\centering
  				\includegraphics[width=1\linewidth]{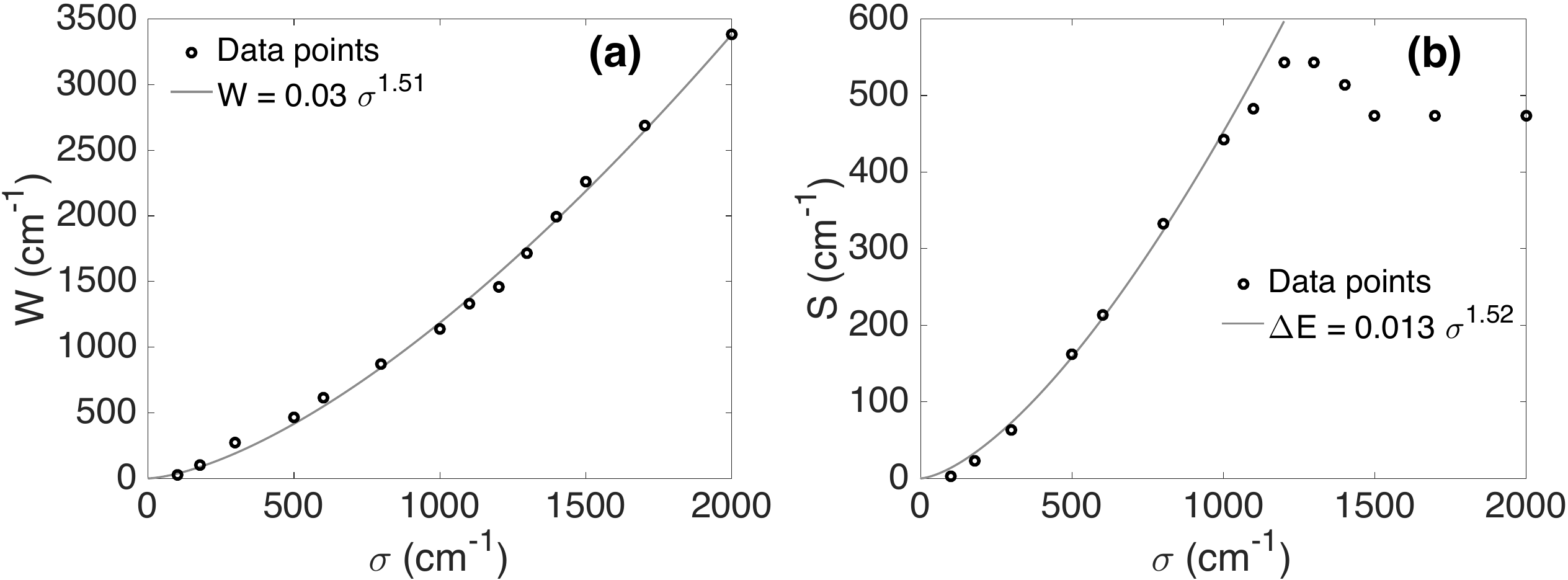}
				\caption{Disorder scaling of the absorption properties of the tube aggregate:
				(a) FWHM, or W, and (b) red shift, or S, of the lowest-energy 
				absorption band of tubes with $N_1=666$ and $N_2=6$.
				\label{Deloc:fig:abs_fwhm}}
				\end{figure}

\textbf{Disorder scaling of absorption band width and position.}
Next, we examine the disorder dependence of the optical band width, 
$W$, and red shift, $S$, of the lowest-energy \textit{J} band. To this end, 
we first fit the absorption spectrum to a sum of Gaussian line shapes in 
order to isolate this \textit{J} band. Then, we take the full width at half 
maximum of the corresponding Gaussian as $W$. For $S$, we use 
the difference between the mean value of the corresponding Gaussian 
and the energy position of the lowest-energy peak in the stick spectrum.
The obtained results for $W$ and $S$ are presented in 
Figs.~\ref{Deloc:fig:abs_fwhm}(a) and \ref{Deloc:fig:abs_fwhm}(b), 
respectively, as a function of the disorder strength.
Both dependencies may be fitted well by a power law 
(curves in Fig.~\ref{Deloc:fig:abs_fwhm}), as is common 
for a variety of molecular aggregates \cite{Schreiber.Toyozawa.I.1982, 
Boukahil.Huber.1990, Kohler.lineshape.1989, Fidder.JCP.1991, 
Malyshev.Moreno.1995, Bloemsma.PRL.2015}.
For the width, the best fit according to $W=a \sigma^b$ yields $b=1.51$ 
(Fig.~\ref{Deloc:fig:abs_fwhm}(a)). The obtained exponent is higher 
than the value of 1.34 obtained for one-dimensional \textit{J} aggregates
\cite{Fidder.JCP.1991,Kohler.lineshape.1989,Schreiber.Toyozawa.I.1982,
Boukahil.Huber.1990}. However, it is considerably smaller than the value 
of 2.83 obtained from a previous study on tubular aggregates~\cite{Bloemsma.PRL.2015}. 
The strong difference with Ref.~\citenum{Bloemsma.PRL.2015} can 
be explained from differences in the exciton density of states at the 
position of the lowest-energy \textit{J} band, which in turn can be traced 
back to differences in the lattice structure. In the case of 
Ref.~\citenum{Bloemsma.PRL.2015}, a tube with one molecule per 
unit cell was considered with a lattice structure which near the lower 
band-edge gives rise to a low density of states, scaling with the square-root 
of energy; in our case, the density of states is rather high already at the position 
of the lowest-energy band and does not seem to depend strongly on energy 
(Fig.~\ref{Deloc:fig:typ_abs}(b)). The result is a scaling of $W$ with 
disorder that is much closer to the one-dimensional case.

The results for the scaling of the energy shift with disorder (Fig.~\ref{Deloc:fig:abs_fwhm}(b)) 
show similar behavior as the absorption band width. The value of 1.52 for 
the exponent in the corresponding power law fit is somewhat larger than 
the value of 1.35 found for the one-dimensional system \cite{Fidder.JCP.1991}.
The increase of the red shift reaches a maximum at  $\sigma=1200$~cm$^{-1}$, 
and then starts to decrease again. This may be explained from the fact that for 
disorder values larger than the exciton bandwidth, the interactions are relatively 
unimportant and the spectrum should tend to a very broad peak, centered at 
the monomer transition frequency.

\subsection{Degree of localization: participation number} 
\label{Deloc:sec:PN}

\begin{figure}[ht]
  				\centering
  				\includegraphics[width=0.6\linewidth]{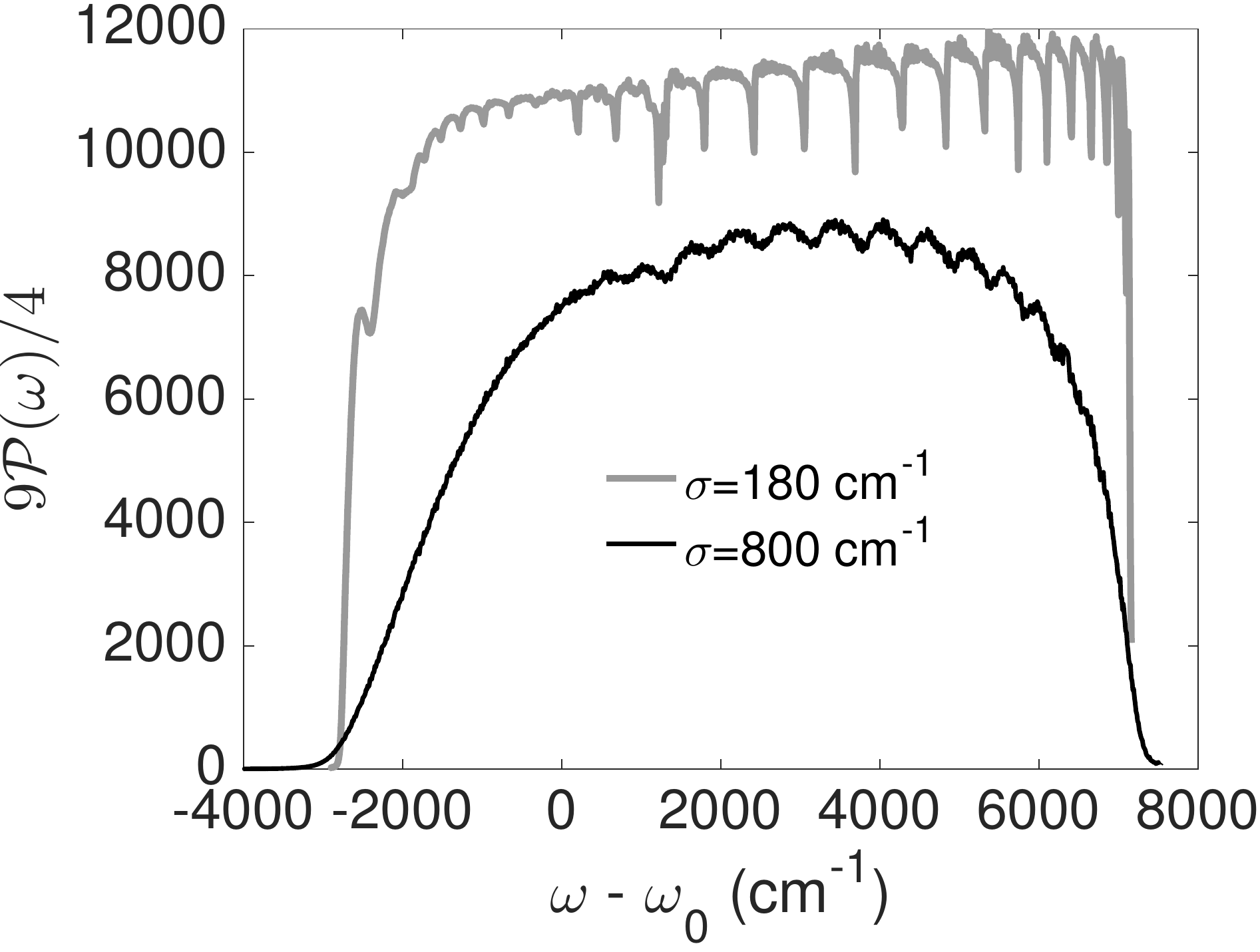}
				\caption{The participation number over the whole exciton 
				band for tubes with $N_2=11$ and $N_1=666$ for two 
				values of the disorder. The scaling factor $9/4$ was 
				introduced to recover the system size in case $\sigma=0$ (see text).
				\label{Deloc:fig:PN_band}}
				\end{figure}

In this section, we establish the behavior of the degree of localization 
of the eigenstates obtained from the participation number calculated 
using Eq.~(\ref{Deloc:eq:PN}). 
The energy dependence of this quantity multiplied by $9/4$ is shown in 
Fig.~\ref{Deloc:fig:PN_band} for $\sigma = 180$~cm$^{-1}$ and 
$\sigma = 800$~cm$^{-1}$, where the factor $9/4$ was introduced 
to ensure that in the homogeneous limit ($\sigma = 0$~cm$^{-1}$), 
this number tends to the system size $2 N_1 N_2$ \cite{Didraga.JCP.2004} 
(the factor of 2 stems from the fact that we deal with tubes with 
2 molecules per unit cell). Clearly, with growing disorder, for each 
energy the states become more localized; furthermore, 
the localization is stronger near the band edges than at 
the band center, as is common for disordered systems
\cite{Anderson.PhysRev.1958, Fidder.JCP.1991}.
Inside the exciton band, most clearly for $\sigma =180$~cm$^{-1}$, 
the participation number exhibits a structure with dips occurring at 
discrete energy positions. This is similar to the dip found in 
one-dimensional disordered systems \cite{Fidder.JCP.1991}
and reflects the persistence of the quasi-one dimensional exciton 
sub-bands characterized by the quantum number $k_2$ for weak 
disorder. For $\sigma=180$~cm$^{-1}$, the value of the participation 
number (scaled by the factor $9/4$) inside the exciton band indeed 
reflects the participation of almost all 14,652 molecules in the exciton 
wave functions, i.e., practically complete delocalization.

Next, we investigate the dependence of the participation number 
on the tube's size (length and radius). We are mainly interested in 
the eigenstates in the optically relevant region of the low-energy 
absorption band, where localization effects are strong. To this end, 
we calculate the average participation number of the exciton states 
in the region of $\pm 80$~cm$^{-1}$ around the peak position of the 
lowest-energy \textit{J} band of the homogeneous aggregate, 
$\approx 16,280$ cm$^{-1}$ (the exact numbers for each system 
are given in the Appendix, Table~\ref{Deloc:tab:L-dep} 
and~\ref{Deloc:tab:R-dep}) denoted as $\mathcal{P}(\omega_J)$.

\begin{figure}[h]
  				\centering
  				\includegraphics[width=0.85\linewidth]{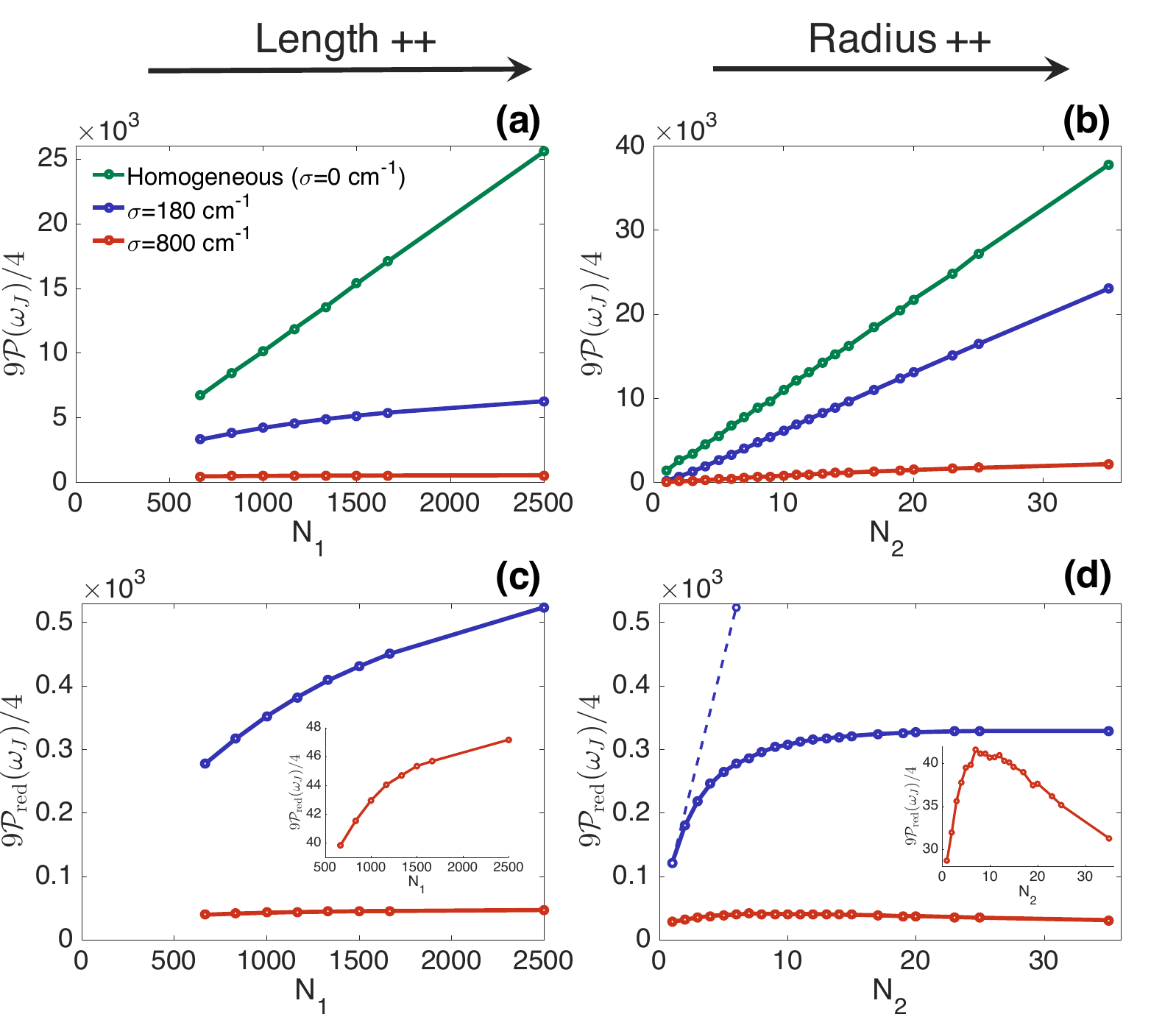}
				\caption{Length and radius dependence of the degree of exciton localization. 
				The participation number (a,b) and the reduced 
				participation number (c,d) near the peak position 
				of the lowest-energy \textit{J} band are shown as a function 
				of system size for two values of the disorder strength: 
				$\sigma=180$~cm$^{-1}$ and $\sigma=800$~cm$^{-1}$. 
				The length dependence (left panels) is shown for tubes 
				with a fixed radius ($N_2=6$) and the length varying from 
				$N_1=666$ to $N_1=2500$. The radius dependence (right 
				panels) is presented for tubes with a fixed length ($N_1=666$) 
				and the radius varying from $N_2=1$ to $N_2=35$. 
				The insets in the bottom panels show blow-ups of the dependence 
				for $\sigma = 800$~cm$^{-1}$. The additional data point in the bottom-right 
				panel for $\sigma=180$~cm$^{-1}$ (connected by the dashed line) 
				indicates the value for $N_2=6$ for a longer tube $(N_1=2500)$. 
				\label{Deloc:fig:ipr}}
				\end{figure}

Fig.~\ref{Deloc:fig:ipr} (top panels) shows $9 \mathcal{P}(\omega_J)/4$ 
as a function of the tube's length and radius for the homogeneous 
system (green) and for two values of the disorder: $\sigma=180$~cm$^{-1}$ 
(blue) and $\sigma=800$~cm$^{-1}$ (red). For the length dependence, 
the radius of the tubes is fixed at $N_2=6$ and the length increases from 
$N_1=666$ to $N_1=2500$. In the case of the radius dependence, the length 
is fixed at $N_1=666$ and the radius increases from $N_2=1$ to $N_2=35$. 
In the homogeneous limit, the participation number (corrected by the factor 
$9/4$) correctly is seen to grow linearly with the system size and to be basically 
equal to this size (in this case $12 N_1$), reflecting complete delocalization. 
Disorder suppresses the exciton delocalization and, therefore, decreases the 
participation number. 
For $\sigma=180$ cm$^{-1}$, the participation number still increases with 
system size over the entire region of the length and radius considered, meaning 
that the delocalization size, even for the largest sizes considered, still is limited 
by the system size and not by the disorder.  
This clearly reflects the weak character of the exciton localization due to the 
higher-dimensional character of the tubes. The radius dependence persists 
longer than the length dependence, which for $N_2 = 6$ starts to saturate 
around a length of 1500 rings. This also is seen for the stronger value of 
the disorder, $\sigma = 800$ cm$^{-1}$, where the length dependence is 
quite weak for $N_2 = 6$, implying that the delocalization of the exciton 
states along the tube's axis is not limited by its length for this disorder strength, 
while the delocalization size grows for growing radius over the entire $N_2$ 
domain studied.

The strong radius dependence of the delocalization sizes 
prompted us to introduce the reduced participation number, 
defined by the participation number divided by the number 
of molecules per ring: $\mathcal{P}_\mathrm{red} = \mathcal{P}/(2N_2)$. 
For states that are completely delocalized around the circumference 
of the tube, $\mathcal{P}_\mathrm{red}$ is expected to be constant 
as a function of $N_2$. This number may then be interpreted as 
the number of rings along the tube over which the exciton states 
are delocalized. The reduced participation number as a function 
of length and radius is shown in the bottom panels of Fig.~\ref{Deloc:fig:ipr}, 
again for $\sigma = 180$~cm$^{-1}$ and 800~cm$^{-1}$. We first 
discuss the data for $\sigma = 180$~cm$^{-1}$. It is clearly seen 
that for small $N_2$ values, up to about $N_2 = 6$, 
$\mathcal{P}_\mathrm{red}$ is not a constant, but grows strongly 
with $N_2$. This means that not only are the states fully delocalized 
around the rings, but that, moreover, the number of rings over which the states 
are delocalized grows with increasing radius.  
This supralinear dependence of the total participation number on 
the radius finds its origin in intra-ring exchange narrowing of 
the disorder: states that are completely delocalized around 
each ring, have $k_2$ states whose energy distribution imposed 
by the disorder does not have a width given by $\sigma$, but rather 
by $\sigma/\sqrt{2N_2}$. In a perturbative picture, for each $k_2$ 
value this leads to an effective one-dimensional chain (of rings) with 
effective energy disorder strength $\sigma^*=\sigma/\sqrt{2N_2}$, 
i.e., an effective disorder strength that diminishes with growing radius. 
This explains that the exciton delocalization along the tube's axis 
can grow with increasing radius. Using the disorder scaling of 
the delocalization size in linear chains found in Ref.~\citenum{Fidder.JCP.1991}, 
the number of rings that participate in the wave functions is expected 
to scale as $\sigma^{*(-2/3)}  \sim N_2^{1/3}$. The actual scaling 
deduced from the first 4 data points for $\mathcal{P}_\mathrm{red}$ 
for $\sigma = 180$~cm$^{-1}$ are best fit to a scaling relation $\sim N_2^{1/2}$. 
Given the difficulty to deduce a good power-law fit from just 4 data 
points and the fact that the perturbative arguments used here are 
bound to break down quite easily for the high density of states in 
the system considered, the differences of the two exponents is not 
unreasonable. Beyond $N_2 \approx 6$, $\mathcal{P}_\mathrm{red}$ 
for $\sigma = 180$~cm$^{-1}$ starts to saturate towards a constant: 
the number of rings that participate in the exciton wave functions hardly 
grows anymore. Closer inspection shows that this saturation is governed 
by the tube's length, i.e., the delocalization size for $N_2 > 6$ is strongly 
limited by the tube's length. This is made explicit by the additional data 
point in the lower-right panel of Fig.~\ref{Deloc:fig:ipr}, which indicates 
$\mathcal{P}_\mathrm{red}$ for $N_2=6$ and $N_1= 2500$.

For the stronger disorder value considered ($\sigma = 800$~cm$^{-1}$), 
intra-ring exchange narrowing also seems to occur for small radii, but 
much less pronounced than for the case of weak disorder (see inset in 
lower-right panel of Fig.~\ref{Deloc:fig:ipr} for details). Moreover, 
following the saturation around $N_2 = 8$, $\mathcal{P}_\mathrm{red}$ 
starts to diminish with growing radius, implying that the states are no 
longer fully delocalized around the tube's circumference. The delocalization 
for this disorder strength is not limited by the chain length of $N_1 = 666$ rings, 
not even for the largest radii.

	\subsection{Extent of the wave function from its autocorrelation function}
	\label{Deloc:sec:ACF}

As mentioned in Sec.~\ref{Deloc:sec:Model}, an alternative 
measure of the degree of delocalization is the auto-correlation 
function of the exciton wave function, which has the advantage that 
for higher-dimensional systems is also gives directional information. 
In this section, we use the auto-correlation function defined in 
Eq.~(\ref{Deloc:eq:CF}) and we will be particularly interested 
in its dependence along the direction of the tube's axis, i.e., in 
$C(s_1, s_2=0; \omega)$ as a function of the relative separation 
$s_1$ between two rings. As before, we will be particularly interested 
in the energy region $\omega_J$ around the lowest-energy 
\textit{J} band.

\begin{figure}[ht]
  				\centering
  				\includegraphics[width=1\linewidth]{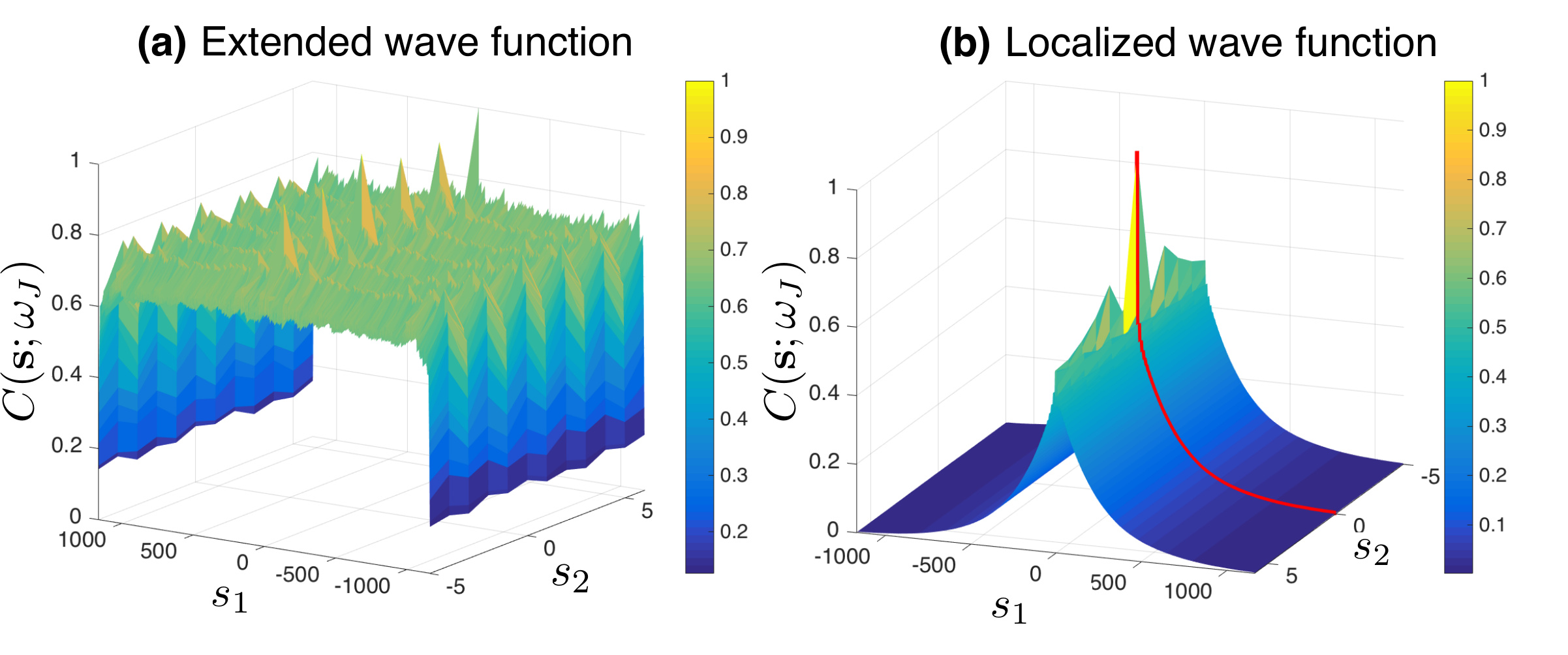}
				\caption{Typical autocorrelation function of the exciton wave function.
				Plotted is $C(\mathbf{s}; \omega_J)$ for a 
				(a)~homogeneous ($\sigma=0$~cm$^{-1}$) and 
				(b) disordered tube ($\sigma=800$~cm$^{-1}$) 
				with $N_1$=1166 and $N_2$=6. The correlation function 
				$C(s_1,s_2=0;\omega)$ is highlighted by the red line.
				\label{Deloc:fig:ACF}}
				\end{figure}

Fig.~\ref{Deloc:fig:ACF} shows the typical autocorrelation 
function for a homogeneous tube with $N_1=1166$ and $N_2=6$ 
(Fig.~\ref{Deloc:fig:ACF}(a)), and the same tube in the presence 
of strong disorder $\sigma=800$~cm$^{-1}$ (Fig.~\ref{Deloc:fig:ACF}(b)). 
As can be seen from the 3D correlation plot, the exciton wave 
function of the homogeneous system (Fig.~\ref{Deloc:fig:ACF}(a)) 
is extended over the whole aggregate, with a steep drop of the correlation 
function at the edges of the tube due to the open boundary conditions. 
In the $s_2$ direction, such drop does not occur, because of the circular 
nature of this coordinate.
For the disordered system (Fig.~\ref{Deloc:fig:ACF}(b)), the autocorrelation 
function at $\omega = \omega_J$ shows a peak with maximum value 1 
at the origin, $(s_1,s_2)-(0,0)$, and a drop in both $s_1$ (longitudinal) 
and $s_2$ (circumferential) directions. 
It can be seen, though, that a high correlation along $s_2$ is preserved, 
which supports the idea that the states at $\omega_J$ still are quite 
strongly delocalized along the rings for this circumference, even for 
$\sigma = 800$~cm$^{-1}$. 
As we are particularly interested in the direction of the tube's axis, 
we study the decay of the correlation function $C(s_1,s_2=0;\omega)$, 
highlighted by the red line in Fig.~\ref{Deloc:fig:ACF}(b).
Initially, the correlation function follows a power-law decay, which only 
is important for a few rings close to the origin, while at larger distances 
the decay is exponential. 
This is in agreement with a previous numerical study of 1D, 2D, and 3D 
disordered electronic systems \cite{Schreiber.1985.JPhysC.18.2493}, 
where it was concluded that a power-law decay of the wave functions 
mediates between extended states and strongly localized states with 
exponential decay.

\begin{figure}[ht]
  				\centering
  				\includegraphics[width=1\linewidth]{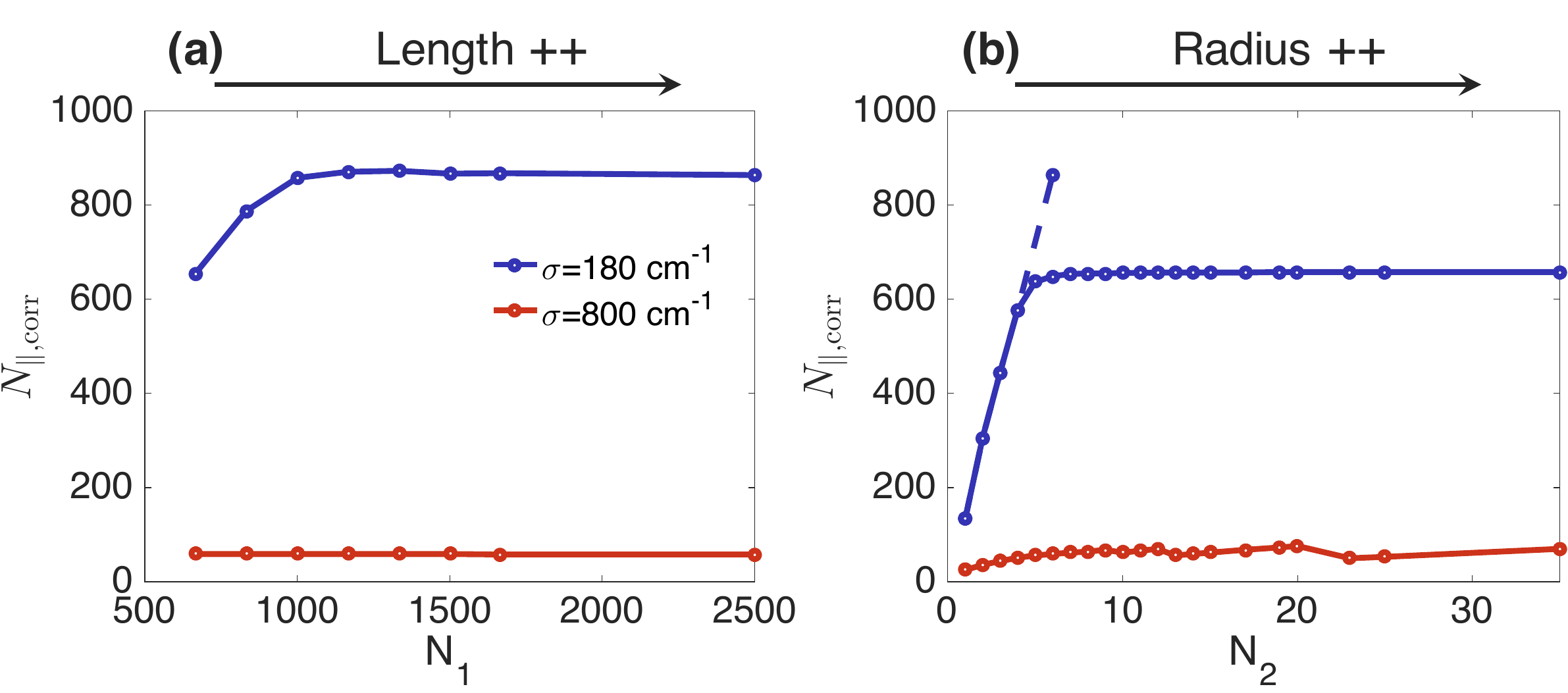}
				\caption{Correlation length as a function of the tube's length and radius. 
				(a) The length dependence is calculated for 
				tubes with a fixed radius ($N_2=6$) and a length 
				varying from $N_1=666$ to $N_1=2500$. 
				(b) The radius dependence is determined 
				for tubes with a fixed length ($N_1=666$) and a radius 
				varying from $N_2=1$ to $N_2=35$. The additional data 
				point connected to the others by a dashed line, is obtained 
				for $N_1=1000$ and $N_2=6$.
				\label{Deloc:fig:CF}}
				\end{figure}

Fig.~\ref{Deloc:fig:CF} shows the correlation length in the longitudinal 
direction $N_{\parallel, \mathrm{corr}}$ (defined above in Eq. (\ref{Deloc:eq:CF})) 
as a function of the tube's length and radius. 
In Fig.~\ref{Deloc:fig:CF}(a) it is seen that for weak disorder 
($\sigma=180$~cm$^{-1}$), $N_{\parallel, \mathrm{corr}}$ grows with 
the tube's length up to a length of about $N_1=1000$, after which 
it reaches a plateau with $N_{\parallel, \mathrm{corr}} \approx 870$. 
This means that for tubes with $N_2=6$ and shorter than about 1000 
rings, the physical size is the limiting factor for the correlation length. 
This matches the saturation of the increase of the participation 
number seen in Fig.~\ref{Deloc:fig:ipr}(a) around $N_1=1000 - 1500$. 
As can be seen in Fig.~\ref{Deloc:fig:ACF}(a), for $\sigma=800$~cm$^{-1}$ 
the disorder is the limiting factor for $N_{\parallel, \mathrm{corr}}$, 
which again matches the behavior seen in Fig.~\ref{Deloc:fig:ipr}(a).

As is seen in Fig.~\ref{Deloc:fig:CF}(b), for weak disorder, 
$N_{\parallel, \mathrm{corr}}$ grows linearly with the radius 
for small values of $N_2$, while a plateau is observed for 
values of $N_2>5$. This plateau results from a physical 
limitation, namely the length $N_1=666$ of the tubes considered 
in case we study the radius dependence. This is confirmed by 
also calculating the correlation length for $N_1$=1000 and 
$N_2$=6 (additional data point in Fig.~\ref{Deloc:fig:CF}(b) 
connection by a dashed line). The linear dependence of 
the correlation length on $N_2$ appears to persist till $N_2=6$. 
This behavior is in agreement with the fact that the reduced 
participation ratio in Fig.~\ref{Deloc:fig:ipr}(d) initially grows 
with the radius, albeit that there the increase was not found 
to be linear. Again, the increase of the correlation length with 
the increase of the tube's radius results from the intra-ring 
exchange narrowing effect. 
For $\sigma = 800$ cm$^{-1}$, the radius dependence of 
the correlation length closely matches that of the reduced 
participation number found in Fig.~\ref{Deloc:fig:ipr}d.

\subsection{Character of the exciton wave function}
\label{Deloc:sec:fractal}

\begin{figure}[h!]
  				\centering
  				\includegraphics[width=0.9\linewidth]{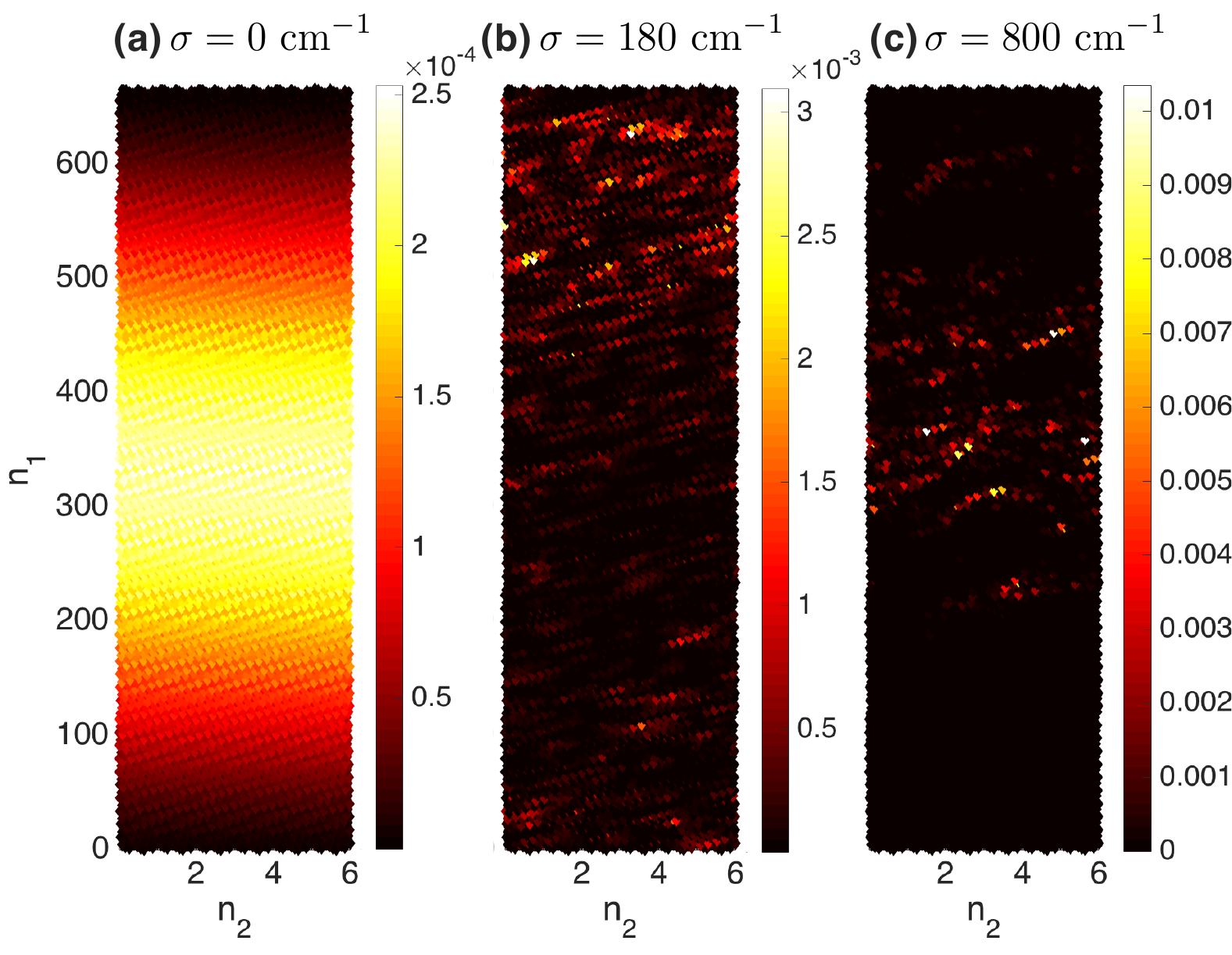}
				\caption{The probability density for wave functions in the low-energy region 
				is plotted on the unwrapped surface for tubes with $N_2=6$ and 
				$N_1=666$: (a) in the absence of disorder, 
				(b)~for $\sigma=180$~cm$^{-1}$, and 
				(c) $\sigma=800$~cm$^{-1}$. 
				\label{Deloc:fig:WF}}
				\end{figure}

The participation number and autocorrelation function give statistical 
information on characteristics of the exciton wave functions. It is also 
interesting, though, to consider examples of typical wave functions for 
a specific realization of the disorder. To this end, we show in Fig.~\ref{Deloc:fig:WF} 
the probability density $|\varphi_{kn}|^2$ on the unwrapped surface of the cylinder 
for typical exciton states near the lowest-energy \textit{J} band for tubes with 
$N_1=666$ and $N_2=6$ with $\sigma = 0$~cm$^{-1}$, 180~cm$^{-1}$, 
and 800~cm$^{-1}$. 
For a homogeneous tube (Fig.~\ref{Deloc:fig:WF}a), the $k_2 = 0$ 
exciton state has equal amplitude on all unit cells within the same ring, 
while along the tube's axis the probability density resembles the lowest exciton 
state in a linear J-aggregate, having a maximum at the center of 
the tube's axis and decaying towards the edges. The alternating pattern, 
observed most clearly in the homogeneous case, is due to the presence 
of two molecules in each unit cell. The chirality observed follows the direction 
of the strongest interaction between neighboring rings.
For weak disorder (Fig.~\ref{Deloc:fig:WF}b), the probability density 
is extended over the whole tube, however, in a quite scattered way, 
similar to the fractal character of quasi-particle states reported 
in disordered two-dimensional systems \cite{Schreiber.1991.multifractal, Aoki.1983}.
For stronger disorder (Fig.~\ref{Deloc:fig:WF}c), the probability density of 
the wave function is more concentrated (localized) on a specific part of 
the cylinder (here the center).
The specific, fractal-like nature of the exciton wave functions at weak disorder 
strengths may be responsible for the large values of both the correlation lengths and the reduced participation numbers 
found for $\sigma = 180$~cm$^{-1}$ described in Sec.~\ref{Deloc:sec:PN} 
and \ref{Deloc:sec:ACF}.

\section{Conclusions} 
\label{Deloc:sec:Conclusions}

In this paper, we systematically examined the dependence 
of the exciton localization and optical properties on both 
the radius and the length of tubular molecular aggregates. 
As a specific model, we used the structure previously reported 
to model the inner wall of C8S3 aggregates, described by 
an extended herringbone model with two molecules per unit cell.
We numerically calculated the absorption spectra in the presence 
of Gaussian diagonal disorder for tubes of various lengths 
(up to 740 nm) and radii (up to 20.7 nm). 
We found that the effect of the tube's length, observed as 
a red-shift of the lowest-energy band with increasing length, 
is still visible for tubes as long as 150 nm. The effect of 
the radius is much more pronounced, due to the strong 
dependence of the higher-energy bands polarized 
perpendicular to the tube's axis on its radius.

We used two quantities to study the localization behavior 
as a function of the length and radius: the (reduced) participation 
number that gives a measure for the typical number of molecules 
participating in the exciton states at a particular energy, and 
the autocorrelation function of the exciton wave function that 
gives statistical information on the extent and the directionality 
of the exciton wave functions.
The obtained results suggest that the physical size rather than 
the disorder is the limiting factor for the delocalization of the exciton 
states of C8S3 cyanine aggregates, at least for aggregates shorter 
than about 1~micron. It should be noted that the length dependence 
of the localization size does not seem to affect the absorption spectrum, 
except for lengths smaller than 150~nm.
In general, we found that for the disorder value relevant to C8S3 
aggregates ($\sigma = 180$ cm$^{-1}$), the exciton wave functions 
in the optically important region of the lowest-energy \textit{J} band 
are fully delocalized around the circumference of the tubes, which 
is consistent with the strong polarization properties found in 
the experimental absorption spectrum of these aggregates \cite{Didraga.JPCB.2004}.
Moreover, this circumferential delocalization persists up to 
large radii, even larger than those considered in our calculations. 
This, inter alia, gives rise to the interesting effect of intra-ring 
exchange narrowing of disorder, which ultimately results in 
the growth of the delocalization length along the axis direction 
of the aggregate with growing radius.
The excitonic states in the middle of the exciton band are hardly 
affected by static disorder, even for strong disorder values of more 
than 1000~cm$^{-1}$. 
Conversely, states at the lower edge of the exciton band (close to 
the lowest-energy \textit{J} band) as well as the upper edge, show 
stronger localization effects than those in the optically dominant 
region, but they still are very delocalized.

The properties of the exciton states found here are of interest in 
their own right, in the realm of localization problems, and for the 
optical absorption
of the system, namely to identify the character of the states that 
are responsible for the absorption process. They may also have 
a bearing on dynamic processes, such as exciton transport, which 
was shown to occur with a higher diffusion constant for larger 
delocalization lengths \cite{Chuang.PRL.2016}. In that case, however, 
a more in-depth study is needed to also assess the importance of 
dynamic disorder, giving rise to dephasing, which over time may 
destroy coherences between different molecules, in particular 
distant ones. 
In addition, we note that in this work, we have restricted ourselves 
to near-field $1/r^3$ dipolar interactions. For the longest aggregates 
considered here, the inclusion of the radiative corrections to these 
interactions may affect our results \cite{Spano1989superradiance,
Celardo.NJP.2019, Celardo.microtubules.2019}, 
which would be interesting to explore further in a future systematic study.

\newpage
\section*{Appendix: Additional information about modeled structures} 

\noindent \textbf{Length distribution.}
For the length dependence studies, the radius of the tubes 
was kept fixed, while the length was varied from 50~nm up to 
740~nm (see Table \ref{Deloc:tab:L-dep}). The radius was 
chosen to be equal to 3.5505 nm, which corresponds to 
$N_2$=6 unit cells in the ring, as is the case for the inner wall 
of C8S3 aggregates \cite{Eisele.NatChem.2012}.

\begin{table}[ht]
\caption{Model tubes used for length dependence study.
$N_1$ is the number of the rings in the tube, $L$ is the length 
of the tube, $N$ is the total number of molecules,
$\omega$ ($k_2=0$) is the frequency of the low-energy 
Davydov component of the $k_2$ = 0 band, and the bandwidth 
refers to the width of the exciton band of the tube without disorder.}
\begin{center}
\begin{tabular}{  c | c | c | c | c  }
\hline\hline 
$N_1$ & $L$, nm & $N$ & Bandwidth, cm$^{-1}$ & $\omega$ ($k_2=0$), cm$^{-1}$
\tabularnewline
\hline
170		&	50.3			& 		2040		&	16,113.2 - 25,991.8	&  16,339	\\
340		&	100.5		& 		4080		&	16,111.8 - 25,995.0	&  16,299	\\
510		&	150.8		& 		6120		&	16,111.5 - 25,995.6	&  16,288	\\
666		&	196.9		& 		7992		&	16,111.4 - 25,995.8	&  16,283	\\
833		&	246.2		& 		9996		&	16,111.4 - 25,995.9	&	16,281	\\
1000	&	295.6		& 		12,000		&	16,111.3 - 25,995.9	&	16,279	\\
1166		&	344.7		&		13,992		&	16,111.3 - 25,996.0	&	16,278	\\
1333	&	394.0		& 		15,996		&	16,111.3 - 25,996.0	&	16,277	\\
1500	&	443.4		&		18,000		&	16,111.3 - 25,996.0	&	16,277	\\
1666	&	492.5		&		19,992		&	16,111.3 - 25,996.0	&	16,276	\\
2500	&	739.0		&		30,000		&	16,111.3 - 25,996.0 	&	16,275	\\
\hline\hline
\end{tabular}
\end{center}
\label{Deloc:tab:L-dep}
\end{table}


\noindent \textbf{Radius distribution.}
For the modeling of the radius dependence, the same lattice 
of the inner wall of C8S3 was used, where the radius 
was defined by the number of unit cells on the rolling vector 
(in order to preserve the rolling angle, only specific radii can 
be taken, namely when the end point of the rolling vector 
coincides with a lattice point). The  radii considered are given 
in Table~\ref{Deloc:tab:R-dep}. The length of the tubes was 
then kept fixed at 196.9~nm.

As mentioned in the main text, the radius of the tube with $N_2=6$ corresponds 
to the inner wall of C8S3 aggregate, while the one with $N_2=11$ agrees 
with the inner wall of bromine-substituted C8S3 aggregate (without adjusting the other structural 
parameters) \cite{Eisele.NatChem.2012, Kriete.JPCL.2017}.

\begin{table}[ht]
\caption{Model tubes used for the radius dependence study. 
$N_2$ is the number of unit cells in the ring, $R$ is the radius, and 
$N$ is the total number of molecules. The bandwidth and $\omega$ ($k_2=0$) 
are as defined in the caption of Table~\ref{Deloc:tab:L-dep}.}
\begin{center}
\begin{tabular}{ c | c | c | c | c  }
\hline\hline 
$N_2$ & $R$, nm & $N$ & Bandwidth, cm$^{-1}$ & $\omega$ ($k_2=0$), cm$^{-1}$
\tabularnewline
\hline
1		&		0.5918		&		1332		&		15,950.2 - 24,885.7		& 16,355	\\
2		& 		1.1835		&		2664		&		16,080.7 - 25,900.1		&	16,273	\\
3		& 		1.7753		&		3996		&		16,103.4 - 25,878.2		&	16,276	\\
4		& 		2.3670		&		5328		&		16,109.1 - 25,980.4		& 16,279	\\
5		& 		2.9588		&		6660		&		16,110.6 - 25,961.4		& 16,281	\\
6		& 		3.5505		&		7992		&		16,111.4 - 25,995.8		& 16,283	\\
7		&		4.1423		&		9324		&		16,112.0 - 25,984.3		& 16,284	\\
8		& 		4.7340		&		10,656		&		16,112.4 - 26,001.2		&	16,286	\\
9		&		5.3258		&		11,988		&		16,112.8 - 25,993.8		&	16,287	\\
10	&		5.9175 	&		13,320		&		16,113.0 - 26,003.7		&	16,287	\\
11		&		6.5093		&		14,652		&		16,113.3 - 25,998.6		&	16,288	\\
12	&		7.1011		&		15,984		&		16,113.4 - 26,005.1		&	16,289	\\
13	&		7.6928		&		17,316		&		16,113.6 - 26,001.3		&	16,290	\\
14	&		8.2846		&		18,648		&		16,113.7 - 26,005.9		&	16,290	\\
15	&		8.8763		&		19,980		&		16,113.8 - 26,003.0		&	16,291	\\
17	&		10.06		&		22,644		&		16,114.0 - 26,004.2		&	16,292	\\
19	&		11.243		&		25,308		&		16,114.1 - 26,005.0		&	16,293	\\
20	&		11.835		&		26,640		&		16,114.2 - 26,007.1		&	16,293	\\
23	&		13.61		&		30,636		&		16,114.3 - 26,006.0		&	16,294	\\
25	&		14.794		&		33,300		&		16,114.3 - 26,006.4		&	16,295	\\
35	&		20.711		&		46,620		&		16,114.4 - 26,007.3		&	16,297	\\
\hline\hline
\end{tabular}
\end{center}
\label{Deloc:tab:R-dep}
\end{table}


\begin{acknowledgments}
We gratefully acknowledge discussions with M. S. Pshenichnikov and B. Kriete.
\end{acknowledgments}


%

\end{document}